\documentclass[a4paper,12pt]{article}    

\usepackage{amsmath}
\usepackage[final]{graphicx}
\newcommand{\be}{\begin{equation}}
\newcommand{\ee}{\end{equation}}
\usepackage[brazil]{babel}
\usepackage[latin1]{inputenc}
\usepackage[T1]{fontenc}

\title{A din\^{a}mica de um elipsoide em rota\c{c}\~{a}o}
\author{Laysa G. Martins\footnote{laysamartinsymail@yahoo.com.br}\\ Jos\'{e} A. C. Nogales\footnote{jnogales@dex.ufla.br} \\
Departamento de Ci\^{e}ncias Exatas, \\
Universidade Federal de Lavras, Campus da UFLA , \\ Caixa Postal 3037 - CEP 37200-000 - Lavras MG - Brazil}

\begin{document}

\maketitle

\begin{abstract}
Um fen\^omeno f\'isico interessante que contraria nosso senso comum
concerne \`a din\^amica do movimento de um elipsoide em rota\c{c}\~ao numa superf\'icie n\~ao lisa.
Um elipsoide em rota\c{c}\~ao \'e, por exemplo, um ovo cozido, girando numa mesa de superf\'icie \'aspera.
Neste artigo, apresentamos uma explica\c{c}\~ao te\'orica, da din\^amica do movimento desse elipsoide,
que descreve o fen\^omeno de eleva\c{c}\~ao do eixo horizontal para a vertical.
As equa\c{c}\~oes de movimento foram estabelecidas utilizando o formalismo lagrangiano. \\

Palavras chaves: mec\^{a}nica anal\'{i}tica, rota\c{c}\~{a}o de um corpo r\'{i}gido, experimento caseiro.
\end{abstract}

\begin{abstract}
An interesting physical phenomenon, which contradicts our common sense,
is concerned with the dynamics of motion of a spinning ellipsoid in a non smooth surface.
A hard-boiled egg spinning on a table with a rough surface is an example.
In  this article, we present a theoretical explanation, of the dynamics of motion of this ellipsoid,
that describes the axis raising phenomenon, from the horizontal to the vertical.
The equations of motion were obtained with Lagrangian formalism. \\

Keywords: analytical mechanics, rigid-body rotation, homemade experiment.
\end{abstract}

\section{Introdu\c{c}\~{a}o}
\indent A descri\c{c}\~{a}o hist\'{o}rica e qualitativa do movimento de rota\c{c}\~{a}o de um corpo r\'{i}gido, do tipo pi\~{a}o 
``tippe top'', j\'{a} \'{e} estudada h\'a longo tempo. H\'{a} relatos que datam de 1800, quando Sir William Thomson e o professor Hugh 
Blackburn perceberam propriedades semelhantes \`as do movimento do pi\~{a}o ``tippe top'', ao girarem pedras encontradas em praias. 
Entretanto, esse pi\~{a}o ainda n\~{a}o havia sido inventado nessa \'{e}poca \cite{1}. Em 1890, John Perry escreveu um livro, 
``\textit{Spinning Tops and Gyroscopic Motion}'', no qual h\'{a} uma descri\c{c}\~{a}o de um objeto esf\'{e}rico que, quando girado sobre 
uma superf\'{i}cie, tal como uma mesa, ter\'{a} uma eleva\c{c}\~{a}o do centro de massa. Somente em 1892, o pi\~{a}o foi patenteado na 
Alemanha e, curiosamente, os modelos descritos no documento da patente pareciam que n\~{a}o estavam funcionando, sendo verificado mais 
tarde, que os modelos estavam, sim, em pleno funcionamento, somente que a descri\c{c}\~{a}o n\~{a}o havia sido feita corretamente no 
documento. De acordo com ``\textit{Vendsyssel Historical Museum}'' na Dinamarca \cite{2}, o pi\~{a}o foi reinventado em 1949, pelo 
engenheiro  dinamarqu\^{e}s, Werner Ostberg. Desde ent\~{a}o, o pi\~{a}o ganhou grande popularidade em todo o mundo, despertando a 
curiosidade de muitas pessoas, veja a figura 1, motivando assim o estudo desse fen\^{o}meno e a partir disso foram publicados v\'{a}rios 
artigos \cite{1}. Apenas em 2002, Moffatt e Shimomura \cite{3} apresentaram uma explica\c{c}\~{a}o para esse comportamento. No trabalho 
deles as equa\c{c}\~oes de movimento foram obtidas a partir da segunda Lei de Newton.\\

\begin{figure}[!htb]
\centering
\includegraphics[scale=0.28,angle=90]{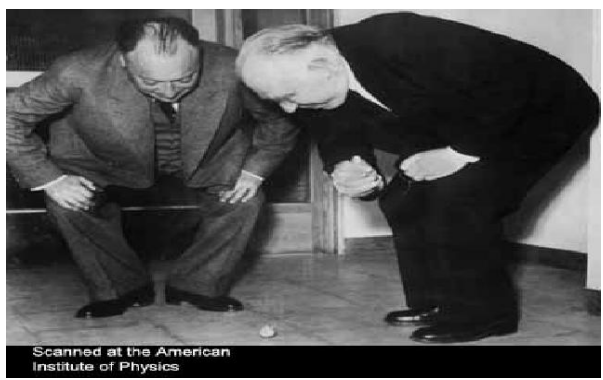} 
\caption{Foto feita na inaugura\c{c}\~{a}o do novo instituto de F\'{i}sica da Universidade de Lund, Su\'{e}cia, em 31 de maio de 1951, 
onde Wolfgang Pauli e Niels Bohr observam o movimento do pi\~{a}o ``tippe top'' \cite{1}.}
\end{figure}

\indent No presente artigo ser\~ao obtidas as equa\c{c}\~{o}es de movimento, da eleva\c{c}\~ao do eixo horizontal para a vertical de um 
elipsoide r\'igido, utilizando as fun\c{c}\~oes de Lagrange, ao inv\'es das leis de Newton como foi feito por Moffatt e Shimomura 
\cite{3}. Na se\c{c}\~{a}o seguinte, ser\'a feita uma descri\c{c}\~ao qualitativa do fen\^omeno f\'isico e posteriormente, faremos uma 
apresenta\c{c}\~{a}o matem\'atica da geometria do problema, impondo considera\c{c}\~{o}es geom\'{e}tricas e defini\c{c}\~{o}es dos 
sistemas de coordenadas que possibilitar\~ao estabelecer a cinem\'{a}tica do sistema. J\'{a} para o caso da din\^{a}mica ser\~ao utilizadas 
as fun\c{c}\~oes de Lagrange, para determinar as equa\c{c}\~{o}es de movimento. Mostraremos que toda a din\^{a}mica pode ser estudada 
somente com uma vari\'{a}vel, a qual descreve o \^{a}ngulo de eleva\c{c}\~{a}o do elipsoide. Ao se  apresentar uma solu\c{c}\~{a}o 
particular, verificaremos a exist\^{e}ncia de um invariante, conhecida como invariante de Jellett, o qual permitir\'a a integra\c{c}\~{a}o 
do sistema. Por meio desse invariante, \'{e} poss\'{i}vel considerar que o sistema corresponde \`a aproxima\c{c}\~{a}o da condi\c{c}\~{a}o 
de equil\'{i}brio do girosc\'{o}pio. Finalmente, faremos an\'{a}lises das equa\c{c}\~{o}es obtidas, envolvendo as dimens\~{o}es do 
elipsoide e a varia\c{c}\~{a}o do seu \^{a}ngulo de eleva\c{c}\~{a}o ao longo de um intervalo de tempo, para diferentes coeficientes de 
atrito. A eleva\c{c}\~{a}o do eixo horizontal para a vertical do elipsoide pode ser vista por meio de um v\'ideo, contendo um experimento 
qualitativo \cite{4}.

\section{Descri\c{c}\~{a}o qualitativa do fen\^{o}meno}
\indent Para a descri\c{c}\~ao qualitativa da eleva\c{c}\~ao do eixo horizontal para a vertical de um elipsoide r\'igido, analisaremos 
o movimento de um ovo cozido que \'{e} girado sobre uma superf\'{i}cie horizontal n\~ao lisa.

\begin{figure}[!htb]
\centering
\includegraphics[scale=0.28,angle=0]{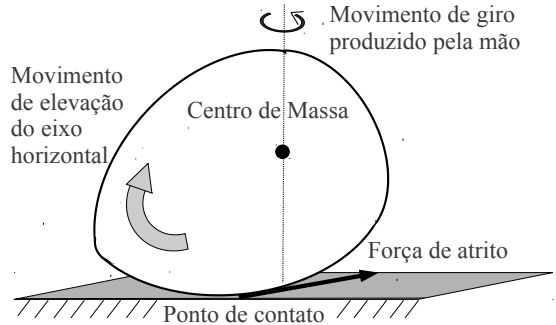} 
\caption{Representa\c{c}\~{a}o do centro de massa, do ponto de contato e da for\c{c}a de atrito no movimento de giro do ovo.}
\end{figure}

\indent Quando um ovo cozido \'e girado sobre uma mesa com uma velocidade suficientemente alta, o seu eixo horizontal eleva-se e passa a 
ocupar a posi\c{c}\~{a}o vertical, veja a figura 2. Em raz\~ao do giro do ovo em torno do seu centro de massa, o ponto de contato \'{e} 
arrastado ao redor da superf\'{i}cie, veja que o centro de massa e o ponto de contato n\~{a}o coincidem. Existe uma for\c{c}a de atrito 
que se contrap\~{o}e ao movimento do ovo, o efeito produzido por essa for\c{c}a de atrito no centro de massa, faz com que surjam 
componentes de torque, os quais atuam tangencialmente \`{a}  superf\'{i}cie do ovo, o que permite a eleva\c{c}\~{a}o do mesmo.

\newpage

\section{Geometria do fen\^{o}meno}
Para encontrar as equa\c{c}\~{o}es de movimento s\~ao consideradas as seguintes a\-pro\-xi\-ma\-\c{c}\~{o}es: o ovo ser\'{a} representado 
por um corpo r\'{i}gido descrito na forma de um elipsoide prolato.\\
\indent S\~{a}o escolhidos dois sistemas de coordenadas, sendo um deles OXYZ fixo no espa\c{c}o e o outro Oxyz fixo no corpo. Esses dois 
sistemas de coordenadas possuem a origem coincidente e esta est\'{a} localizada no centro de massa do corpo. Na figura 3, ilustra-se esta 
descri\c{c}\~{a}o, sendo o eixo Oz sim\'{e}trico.\footnote{Se existe um eixo de simetria em um corpo r\'{i}gido, este ser\'{a} sempre um 
eixo principal, tendo, como propriedade importante, que a dire\c{c}\~{a}o do momento angular \'{e} a mesma que a da velocidade angular.}

\begin{figure}[!htb]
\centering
\includegraphics[scale=0.4,angle=0]{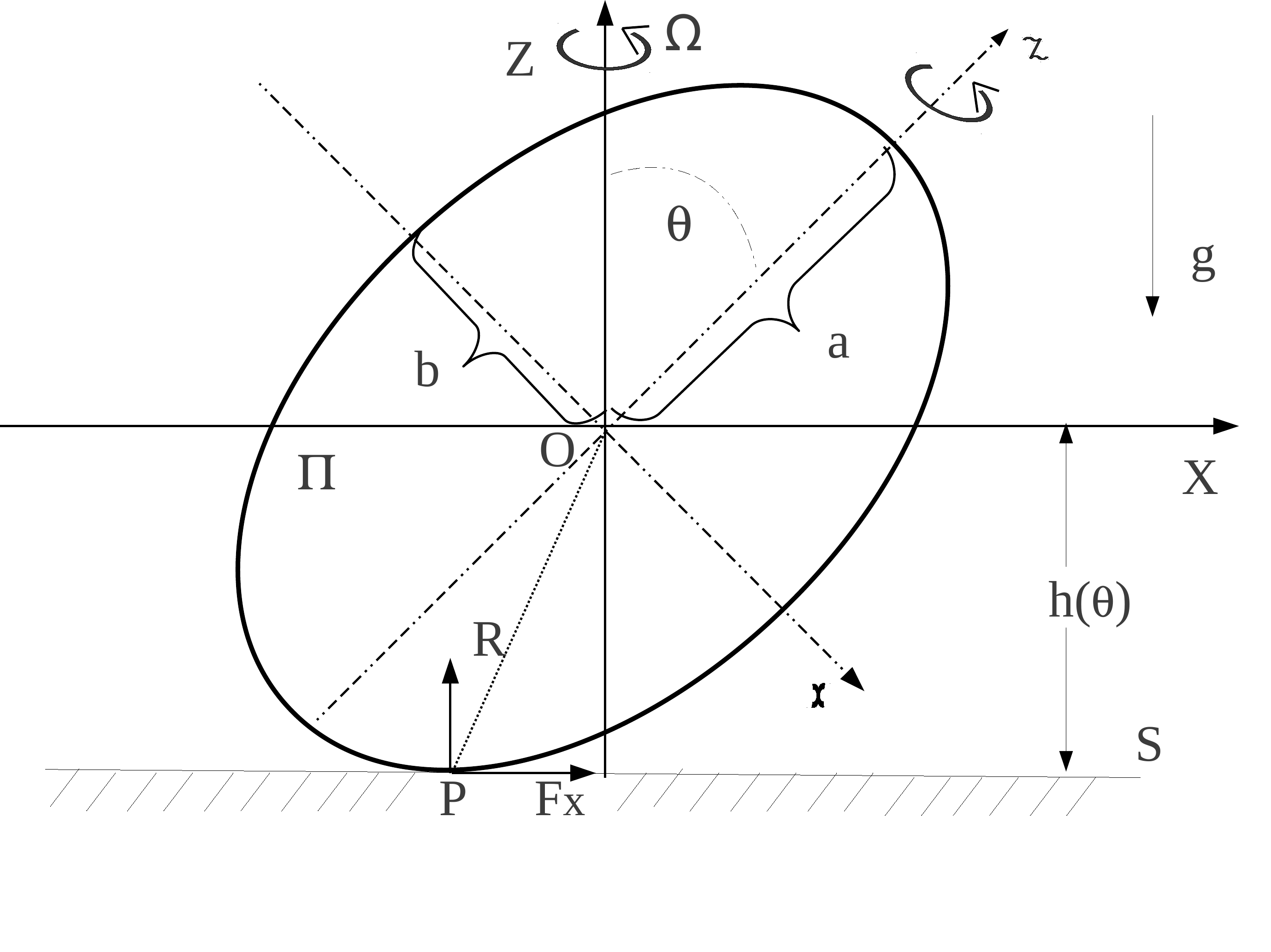} 
\caption{Elipsoide de dimens\~{o}es {\textit{a}} e {\textit{b}} sobre uma superf\'{i}cie S com ponto de contato P. H\'{a} dois sistemas 
de refer\^{e}ncia, OXYZ e Oxyz, sendo o eixo OX horizontal ao plano $\Pi$. A dist\^{a}ncia entre a superf\'{i}cie S e o eixo OX \'{e} dado 
por $h(\theta)$, onde $\theta$ \'e o \^angulo de eleva\c{c}\~ao do elipsoide.}
\end{figure}

\indent Para uma pequena varia\c{c}\~{a}o na orienta\c{c}\~{a}o do corpo em rela\c{c}\~{a}o \`{a} mesa, as coordenadas OX, OY, Oz para o 
ponto P s\~{a}o respectivamente:\footnote{Neste artigo, todas as equa\c{c}\~{o}es est\~{a}o descritas em termos do sistema de coordenada 
OXYZ.}

\be
X_{p}=\frac{dh}{d\theta},
Y_{p}= 0  \quad\mbox{e}\quad 
Z_{p}=-h(\theta).
\ee

\indent Por meio da semelhan\c{c}a de tri\^{a}ngulos ret\^{a}ngulos, pode-se determinar a altura entre a origem O dos sistemas de 
coordenadas com a superf\'{i}e S da mesa, encontrando:

\be
h(\theta)=(a^2\cos^2\theta+b^2\sin^2\theta)^{1/2},
\ee

\noindent onde $a$ e $b$ s\~{a}o as dimens\~{o}es do elipsoide e $\theta$ o \^{a}ngulo que indica o comportamento de eleva\c{c}\~{a}o do 
eixo horizontal (Oz) para a vertical (eixo Oz coincidindo com o eixo OZ).\\
\indent Por meio da aproxima\c{c}\~{a}o desse corpo r\'{i}gido por um elipsoide prolato, t\^{e}m-se que este possui a forma 
$a^2(x^2+y^2)+b^2(z-d)²=a^2b^2$. Para um elipsoide uniforme $d=0$, pois n\~{a}o haver\'{a} deslocamento do centro de massa no sistema OXYZ.\\
\indent Para descrever o movimento de um corpo r\'{i}gido, s\~{a}o necess\'{a}rios seis vari-\'{a}veis. Normalmente, tr\^{e}s dessas 
vari\'{a}veis s\~{a}o tr\^{e}s coordenadas que localizam a posi\c{c}\~{a}o  do centro de massa que, nesse caso, coincidem com a origem do 
sistema de coordenada do corpo e tr\^{e}s \^{a}ngulos independentes que fornecem a orienta\c{c}\~{a}o do sistema de coordenada fixo no 
corpo em rela\c{c}\~{a}o ao sistema de coordenada fixo no espa\c{c}o. Esses \^{a}ngulos $(\varphi,\theta,\psi)$ s\~{a}o conhecidos como 
\^{a}ngulos de Euler \cite{5}.\\
\indent Como as origens dos  sistemas de coordenadas coincidem, o vetor posi\c{c}\~{a}o para um ponto qualquer do corpo r\'{i}gido \'{e} 
dado por:

\be
\textbf{x}=x\textbf{i}+y\textbf{j}+z\textbf{k}=X\textbf{I}+Y\textbf{J}+Z\textbf{K},
\ee

\noindent e a rela\c{c}\~{a}o entre os sistemas de coordenadas para a posi\c{c}\~{a}o, s\~{a}o:

\begin{equation}
 X= x\cos\theta+ z \sin\theta \quad\mbox,\quad
 Y= y  \quad\mbox{e}\quad 
 Z= -x\sin\theta+ z \cos\theta.
\end{equation}

\noindent Em termos dos vetores unit\'{a}rios, ($\textbf{i},\textbf{j},\textbf{k}$) e ($\textbf{I},\textbf{J},\textbf{K}$) \'{e} dada por:

\be
{\bf I}={\bf i}\cos\theta+\textbf{k}\sin\theta \quad\mbox,\quad
{\bf J}={\bf j}  \quad\mbox{e}\quad 
{\bf K}=-{\bf i}\sin\theta+{\bf k}\cos\theta.
\ee

\noindent Finalmente, o vetor posi\c{c}\~{a}o ${\bf X{p}}=(X{p},Y{p},Z{p})$ que localiza o ponto de contato P, \'{e}:

\be
X_{p}= x_{p}\cos\theta+ z_{p}\sin\theta \quad\mbox,\quad
Y_{p}= y_{p}  \quad\mbox{e}\quad 
Z_{p}= -x_{p}\sin\theta+ z_{p} \cos\theta.
\ee

\indent Dos \^{a}ngulos de Euler pode-se perceber que $\varphi$ representa a rota\c{c}\~{a}o em torno do eixo OZ, $\theta$ o \^{a}ngulo de 
eleva\c{c}\~{a}o e $\psi$ \'{e} a composi\c{c}\~{a}o da rota\c{c}\~{a}o em torno do eixo OZ e do eixo Oz. A rela\c{c}\~{a}o entre esses 
\^{a}ngulos e o eixo OZ \'{e} dada por:

\be
\Lambda=\dot{\theta}, \quad\mbox\quad 
\Omega=\dot{\varphi}  \quad\mbox{e}\quad 
n=\dot{\psi}+\Omega\cos\theta,
\ee

\noindent onde $\Omega$ \'{e} a taxa de varia\c{c}\~{a}o do plano $\Pi$ em rela\c{c}\~{a}o ao eixo OZ, $\Lambda$ \'{e} a taxa de 
varia\c{c}\~{a}o da inclina\c{c}\~{a}o do eixo de simetria e $n$ \'{e} o spin ao redor do eixo de simetria Oz do corpo. \\
\indent As velocidades angulares dos sitemas de coordenadas OXYZ e Oxyz, s\~{a}o respectivamente,

\be
  {\bf \Omega}=\Omega{\bf K}=-\Omega\sin\theta{\bf i}+\Omega\cos\theta{\bf k} \quad\mbox{e}\quad
  {\bf \omega}= {\bf \Omega}+\Lambda {\bf j},
\ee

\noindent onde ${\bf \omega}$ \'{e} a velocidade angular total no sistema de coordenadas do corpo, que pode ser reescrito na forma de,

\be
  {\bf{\omega}}=((n-\Omega\cos\theta)\sin\theta,\dot\theta,\Omega\sin^2\theta+n\cos\theta),
\ee

\noindent sabendo que $\dot\theta$ \'{e} a diferencia\c{c}\~{a}o feita em rela\c{c}\~{a}o ao tempo. O momento angular \'{e} dado por 
${\bf H}=I{\bf \omega}$, o qual possui as seguintes componentes,

\be
  {\bf H}=((Cn-A\Omega\cos\theta)\sin\theta,A\Lambda,\Omega\sin^2\theta+Cn\cos\theta),
\ee

\noindent onde A e C s\~{a}o os principais momentos de in\'{e}rcia, posto que x e y s\~{a}o sim\'{e}tricos. Esses principais momentos de 
in\'{e}rcia podem ser calculados por meio dos eixos fixados no corpo e que giram com ele.

\section{Equa\c{c}\~{a}o de movimento}
Para estabelecermos as equa\c{c}\~{o}es de movimento, devemos inicialmente encontrar a lagrangiana do sistema, mas como o sistema tem 
for\c{c}a de atrito, precisaremos determinar as for\c{c}as generalizadas, as quais s\~{a}o:
\begin{eqnarray}
 F_{\theta}&=&Z_{p}F_{x}-X_{p}R, \label{11} \\ 
 F_{\psi}&=&x_{p}F_{y}, \label{12} \\ 
 F_{\varphi}&=&-z_{p}F_{y}\sin\theta + x_{p}F_{y}\cos\theta. \label{13}
\end{eqnarray}
\noindent A obten\c{c}\~{a}o dessas equa\c{c}\~{o}es est\'{a} contida no ap\^{e}ndice. Agora podemos es\-ta\-be\-le\-cer as 
equa\c{c}\~{o}es de movimento utilizando as equa\c{c}\~{o}es de Euler-Lagrange. \\ 
\indent Vamos, primeiramente, encontrar a lagrangiana. Considere ${\bf U}=U{\bf I}+V{\bf J}+W{\bf K}$ a velocidade do centro de massa. 
A velocidade no ponto P pode ser encontrada pela equa\c{c}\~{a}o que relaciona a velocidade do centro de massa com a rota\c{c}\~{a}o por:

\be
  {\bf U_{p}}={\bf U}+{\bf \Omega}\times{\bf X_{p}}=U_{p}{\bf I}+V_{p}{\bf J}+W_{p}{\bf K},
\ee

\noindent onde 

\begin{eqnarray}
  U_{p}&=&U-\dot\theta h(\theta) \quad\mbox,\quad \nonumber \\
  V_{p}&=&V+H(\theta)\sin\theta(n-\Omega\cos\theta)+\Omega h'(\theta) \quad\mbox{e}\quad \nonumber \\
  W_{p}&=&W-\dot\theta h'(\theta).
\end{eqnarray}

A lagrangiana \'{e} definida por $L = T - U'$, onde T \'{e} a energia cin\'{e}tica e $U'$ a energia potencial. Sabe-se que a energia 
cin\'{e}tica de um corpo r\'{i}gido pode ser determinada em termos da soma da energia cin\'{e}tica de transla\c{c}\~{a}o (movimento do 
centro de massa) com a energia cin\'{e}tica de rota\c{c}\~{a}o (varia\c{c}\~{a}o angular). Logo, pode-se escrever:

\begin{displaymath}
 L = T - U'= \frac{1}{2}{\bf{U}}\cdotp{\bf{U}} + \frac{1}{2}{\bf{\omega}}\cdotp{\bf{H}}.
\end{displaymath}

\noindent Note que a energia potencial $U'$ \'{e} nula, em decorr\^encia do sistema de coordenada escolhido. A lagrangiana em termos das 
coordenadas generalizadas \'{e}:

\begin{center}
$L = L(X,Y,Z,U,V,W,\theta,\varphi,\psi,\dot\theta,\dot\varphi,\dot\psi). $ 
\end{center}

Para determinar a lagrangiana, primeiro calcula-se:

\begin{center}
${\bf{U}}\cdotp{\bf{U}} = U^2 + V^2 + W^2$ \qquad 
${\bf {\omega}}\cdotp{\bf{H}} = Cn^2 + A \Omega^2\sin^2\theta + A\Lambda^2.$
\end{center}

\noindent Escrevendo ${\bf{\omega}}\cdotp{\bf{H}}$ em termos das coordenadas generalizadas:

\begin{center}
${\bf{\omega}}\cdotp{\bf{H}}= C\dot\psi^2 + 2C\dot\psi\dot\varphi\cos\theta + C\dot\varphi^2\cos^2\theta + A\dot\varphi^2\sin^2\theta + A\dot\theta^2,$ 
\end{center}

\noindent Obtendo assim:

\begin{center}
$ L = \frac{1}{2}(U^2 + V^2 + W^2 + C\dot\psi^2 + 2C\dot\psi\dot\varphi\cos\theta + C\dot\varphi^2\cos^2\theta + A\dot\varphi^2\sin^2\theta + A\dot\theta^2).$
\end{center}

Como, o movimento de giro de um esferoide, com giro suficientemente r\'{a}pido, \'{e} caracterizado pela inexist\^{e}ncia de movimento na 
dire\c{c}\~{a}o $\hat{K}$, em raz\~ao de o ovo permanecer em contato com a superf\'{i}cie, a componente $W_{p}$ dessa velocidade \'{e} nula. 
Portanto, a lagrangiana do sistema fica,

\begin{equation}
L = \frac{1}{2}(U^2 + V^2 + C\dot\psi^2 + 2C\dot\psi\dot\varphi\cos\theta + C\dot\varphi^2\cos^2\theta + A\dot\varphi^2\sin^2\theta + A\dot\theta^2). \label{16}
\end{equation}

Para determinar as equa\c{c}\~{o}es de movimento, utiliza-se a equa\c{c}\~{a}o de Euler-Lagrange:

\be
\frac{d}{dt}\bigg( \frac{\partial L}{\partial \dot q_{i}} \bigg) - \frac{\partial L}{\partial q_{i}} = Q_{i} \label{17}
\ee

\noindent onde $q_{i}$ s\~{a}o as coordenadas generalizadas e $Q_{i}$ s\~{a}o as for\c{c}as generalizadas dadas pelas equa\c{c}\~{o}es 
(\ref{11}), (\ref{12}) e (\ref{13}).\\
\indent As for\c{c}as que atuam no corpo, no ponto P, s\~{a}o a rea\c{c}\~{a}o normal $\textbf{R}=(0,0,R)$, o peso $\textbf{W}=(0,0,-Mg)$ 
e a for\c{c}a de atrito $\textbf{F}=(F,F,0)$, sendo que $\textbf{F}$ est\'{a} em fun\c{c}\~{a}o de $\textbf{U}_{p}$, a qual \'{e} dada 
pela  lei de atrito din\^{a}mico entre duas superf\'{i}cies de contato, assumindo a condi\c{c}\~{a}o 
${\bf F}.{\bf U}_{p}\leq 0$, se ${\bf F}.{\bf U}_{p}=0$ ent\~{a}o ${\bf F}$ e ${\bf U}_{p}$ s\~{a}o ortogonais, onde $\textbf{U}_{p}$ 
\'{e} a velocidade no ponto P.\\
\indent Vamos, agora, determinar as equa\c{c}\~{o}es de movimento, usando (\ref{17}). Seja $\textbf{x}$ o vetor posi\c{c}\~{a}o, ent\~{a}o 
as equa\c{c}\~{o}es de movimento associadas \`as coordenadas cartesianas s\~{a}o:

\begin{displaymath}
\frac{d}{dt} \bigg( \frac{\partial L}{\partial \bf{\dot x}} \bigg) + {\bf \Omega} \wedge {\bf U} = (\dot U - \Omega V,\dot V + \Omega U, 0),
\end{displaymath}

\noindent e como $Q_{i}=(F_{x},F_{y},R)$, ent\~{a}o usando a equa\c{c}\~{a}o (\ref{17}) temos:
\begin{center}
 $(\dot U - \Omega V,\dot V + \Omega U, 0)=(F_{x},F_{y},R).$
\end{center}
\noindent Logo, as equa\c{c}\~{o}es de movimento s\~{a}o:

$$
\left\{
\begin{array}{ccccccc}
 \dot U &=& \Omega V + F_{x}. \\
 \dot V &=& - \Omega U + F_{y} \\
\end{array}
\right.
$$

A equa\c{c}\~{a}o de Lagrange para as componentes angulares, $(\theta,\psi,\varphi)$, s\~{a}o respectivamente:

\begin{displaymath}
\frac{\partial L}{\partial \theta}= - C \dot \psi \dot \varphi\sin\theta - C \dot \varphi ^2\sin\theta\cos\theta + A \dot \varphi ^2 \sin\theta\cos\theta
\end{displaymath}

\begin{displaymath}
\frac{\partial L}{\partial \dot \theta} =A \dot \theta   \quad\mbox,\quad
\frac{d}{dt} \bigg( \frac{\partial L}{\partial \dot \theta} \bigg) = A \ddot \theta
\end{displaymath}

\noindent Substituindo em (\ref{17}), obt\^{e}m-se,

\begin{displaymath}
A \ddot \theta + C \dot \psi \dot \varphi\sin\theta + C \dot \varphi ^2\sin\theta\cos\theta - A \dot \varphi ^2 \sin\theta\cos\theta = Z_{p}F_{x}-X_{p}R.
\end{displaymath}

\noindent Reescrevendo a equa\c{c}\~{a}o acima em termos de $\Lambda$, $\Omega$ e n, teremos:

\begin{displaymath}
A \dot \Lambda + C \Omega\sin\theta (n-\Omega \cos\theta) + C \Omega ^2\sin\theta\cos\theta - A \Omega ^2 \sin\theta\cos\theta = Z_{p}F_{x}-X_{p}R.
\end{displaymath}

\indent Para determinar as equa\c{c}\~{o}es de Lagrange para as componentes $\psi$ e $\varphi$, procede-se analogamente ao processo feito 
para $\theta$. Portanto encontra-se:

\be
\dot \Lambda = \frac{\Omega\sin\theta (A \Omega \cos\theta-Cn) + Z_{p}F_{x}-X_{p}R}{A} \label{18}
\ee

\be
\dot n= \frac{x_{p}F_{y}}{C}
\ee

\be
\dot \Omega = \frac{- 2 \Omega \Lambda \cos\theta + (C/A)n\Lambda + z_{p}F_{y}/A}{\sin\theta} \label{20}
\ee

\noindent Finalmente as equa\c{c}\~{o}es de movimento s\~{a}o:

\begin{eqnarray}
\left\{
\begin{array}{cccccccccccccccc}
  \dot{U} & = & \Omega V+F_{x},\\
\noalign{\medskip} 
  \dot{V} & = & - \Omega U+F_{y},\\
\noalign{\bigskip} 
  \dot{\Omega} & = & [-2\Omega\Lambda\cos\theta + (C/A)n\Lambda +z_{p}F_{y}/A]/\sin\theta,\\
\noalign{\bigskip} 
  \dot{\Lambda} & = & [\Omega\sin\theta(A\Omega\cos\theta-Cn)-RX_{p}+Z_{p}F_{x}]/A,\\
\noalign{\bigskip} 
  \dot{\theta} & =& \Lambda,\\
\noalign{\medskip} 
  \dot{n} & = & x_{p}F_{y}/C.
\end{array}
\right.
\end{eqnarray}

\noindent as quais s\~{a}o as mesmas encontradas por Moffatt e Shimomura, usando as leis de Newton. Dessas equa\c{c}\~{o}es, pode-se 
perceber que, de fato, a for\c{c}a de atrito fornece contribui\c{c}\~{o}es para o movimento, seja este de transla\c{c}\~{a}o 
(equa\c{c}\~{o}es das acelera\c{c}\~{o}es do centro de massa) e de rota\c{c}\~{a}o (equa\c{c}\~{o}es das acelera\c{c}\~{o}es dos 
\^{a}ngulos de Euler). Para as equa\c{c}\~{o}es que envolvem os \^{a}ngulos de Euler, t\^{e}m-se que as componentes da for\c{c}a de 
atrito e da rea\c{c}\~{a}o normal provocam torques, em especial, a acelera\c{c}\~{a}o do \^{a}ngulo $\theta$, a qual determina a rapidez 
da varia\c{c}\~{a}o do eixo horizontal para a vertical, que \'{e} dado pela composi\c{c}\~{a}o dos torques que envolvem essas duas 
for\c{c}as. A seguir, ser\'{a} feita uma an\'{a}lise mais detalhada da contribui\c{c}\~{a}o dada pela for\c{c}a de atrito.

\section{Solu\c{c}\~{a}o Particular}
Nesta se\c{c}\~{a}o, vamos descrever as considera\c{c}\~{o}es feitas por Moffatt e Shimomura. Para a componente $U_{p}$ da velocidade 
tem-se que se $\dot\theta$ for pequeno e se praticamente n\~{a}o houver varia\c{c}\~{a}o do centro de massa, $U=0$.  Logo $U_{p}=0$, para 
qualquer estado de estabilidade e a velocidade no ponto P passa a ser $\textbf{U}_{p}=(0,V_{p},0)$. Como s\'{o} h\'{a} movimento na 
componente-Y, a for\c{c}a de atrito ter\'{a} as seguintes componentes, $\textbf{F}=(0,F,0)$. Consequentemente, a equa\c{c}\~{a}o (\ref{18}) pode ser reescrita, obtendo:

\be
  A\ddot\theta+\Omega\sin\theta(Cn-A\Omega\cos\theta)=-RX_{p}.
\ee

\noindent Al\'{e}m disso, consideramos que se a rapidez da varia\c{c}\~{a}o do \^{a}ngulo $\theta$ em rela\c{c}\~{a}o ao tempo, for 
pequena, $|\ddot\theta|<<\Omega^2$, portanto o termo que cont\'{e}m $\ddot\theta$ pode ser negligenciado e se $\Omega^2$ for considerado 
grande, ent\~{a}o este dominar\'{a} o termo $-RX_{p}$ dado pela equa\c{c}\~{a}o (\ref{18}) resultando em um equa\c{c}\~{a}o de primeira 
ordem, $(Cn-A\Omega\cos\theta)\Omega\sin\theta=0$. Mas, necessariamente, $\sin\theta\neq0$, pois deseja-se estudar o mecanismo que ocorre 
durante a eleva\c{c}\~{a}o do eixo, assim

\be
  Cn=-A\Omega\cos\theta, \label{23}
\ee

\noindent sendo essa a condi\c{c}\~{a}o de equil\'{i}brio do girosc\'{o}pio, na qual as resultantes das for\c{c}as externas e torques 
externos que agem sobre o corpo s\~{a}o nulas. Nessa situa\c{c}\~{a}o, a energia cin\'{e}tica \'{e} bem maior que a energia potencial, 
pois os efeitos produzidos pela for\c{c}a de Coriolis superam os efeitos gerados pela for\c{c}a de atrito e gravitacional, permitindo uma 
aproxima\c{c}\~{a}o para a condi\c{c}\~{a}o de equil\'{i}brio do girosc\'{o}pio, esta encontrada por Moffatt, Shimomura e Branicki [6], a 
partir de $|\ddot\theta|<<\Omega^2$ e portanto, $\Omega^2 \gg [5g\mid a^2-b^2\mid]/[(a^2+ b^2)min(ab)]$.\\
\indent Utilizando a condi\c{c}\~{a}o dada por (\ref{23}), o momento angular passa a ser para as equa\c{c}\~{o}es (\ref{18}) e (\ref{20}), respectivamente,

\begin{center}
$  A\frac{d}{dt}[\Omega\sin\theta]=-FZ_{p}		  \quad\mbox{e}\quad
  C\frac{d}{dt}[(A/C)\Omega\cos\theta]=-FX_{p},$
\end{center}

\noindent por\'{e}m, se a varia\c{c}\~{a}o de $\theta$ for lenta, ent\~{a}o $\theta\approx 0$ e as equa\c{c}\~{o}es acima tornam-se, 
$A\Omega\dot\theta=FZ_{p}$ e $A\dot\Omega=FX_{p}$, se esse resultado for integrado em ambos os lados, obter-se-\'a $A\Omega h(\theta)=$ 
constante. Essa constante n\~{a}o depende da natureza da for\c{c}a de atrito e \'{e} conhecida como invariante de Jellett. Tal constante 
foi encontrada anteriormente e \'{e}:

\be
  J=-{\bf H}.{\bf X}_{p}=A\Omega h. \label{24}
\ee

\noindent Mas, para satisfazer a condi\c{c}\~{a}o (\ref{23}) de equil\'{i}brio do girosc\'{o}pio $dJ/dt=0$ e assim $J=$ constante.

\begin{eqnarray}
  \frac{dJ}{dt} &=& (A\Omega\cos\theta-Cn)(h''\sin\theta-h'\cos\theta)\dot\theta \nonumber \\
               &=& (Cn-A\Omega\cos\theta)h'^2\frac{d}{dt}  \left( \frac{\sin\theta}{h'(\theta)} 
\right). \label{25}
\end{eqnarray}

\indent  Por outro lado, o estado de estabilidade a velocidade V do centro de massa, para o caso do esfer\'{o}ide \'{e} nula, ent\~{a}o a 
velocidade $V_{p}$ torna-se,

\be
  V_{p}=(\Omega\sin^2\theta +n\cos\theta)dh/d\theta +(n-\Omega\cos\theta)h(\theta)\sin\theta. \label{26}
\ee

\indent Utilizando as equa\c{c}\~{o}es (\ref{23}), (\ref{24}) e (\ref{26}) e fazendo $\zeta (\theta)$ $=$ $(\sin^2\theta +(A/C)\cos^2\theta)^{-1/2}$ a velocidade $V_{p}$ pode ser reescrita como,

\be
  V_{p}=(J/A)\zeta^{-3}h^{-1}d(\zeta h)/d\theta.
\ee

\noindent Substituindo as equa\c{c}\~{o}es (1) e (\ref{25}) na equa\c{c}\~{a}o $A\Omega\dot\theta=FZ_{p}$, obt\'{e}m -se:

\be
  J\dot\theta=-Fh^2(\theta). \label{28}
\ee

\noindent Esta \'{e} a solu\c{c}\~{a}o din\^{a}mica do sistema analisado. Agora, precisamos estabelecer a for\c{c}a de atrito para integrar 
o sistema. Para isso, foram consideradas duas alternativas para a for\c{c}a de atrito. A primeira, quando esta admite a lei de Coulomb 
\cite{7,8}, $F=-\mu MgV_{p}/|V_{p}|$, obtendo a solu\c{c}\~{a}o para (\ref{28}),

\begin{equation}
\tan\theta=(a/b)\tan \mu q(t-t_{0}), \label{29}
\end{equation}

\noindent onde $q=Mgab(a-b)/|a-b||J|$ e $t_{0}$ \'{e} uma constante de integra\c{c}\~{a}o. Para o intervalo de tempo finito, 
$\Delta t=\pi /2|\mu q|$, a solu\c{c}\~{a}o (\ref{29}) representa a transi\c{c}\~{a}o do estado inst\'{a}vel para o estado est\'{a}vel. 
No entanto, h\'{a} regi\~{o}es onde $V_{p}$ \'{e} zero, nos quais a equa\c{c}\~{a}o (\ref{29}) n\~{a}o \'{e} anal\'{i}tica, 
impossibilitando uma an\'{a}lise para estados estacion\'{a}rios.\\
\indent A segunda alternativa \'{e} quando a for\c{c}a de atrito assume a lei de atrito viscoso \cite{7,8}, dado por $F=-\mu MgV_{p}$, 
que, quando integrado, resulta em,

\begin{equation}
 \tan\theta=e^{-\mu q(t-t_{0})}, \label{30}
\end{equation}

\noindent onde $q=5g(a^2-b^2)/2(a^2+b^2)$. A rela\c{c}\~{a}o (\ref{30}) mostra que o \^{a}ngulo $\dot\theta$ depende das dimens\~{o}es do 
elipsoide, da gravidade e do coeficiente de atrito.\\
\indent A seguir, faremos a an\'{a}lise num\'{e}rica da solu\c{c}\~{a}o para as duas alternativas e observaremos o comportamento do 
elipsoide. Considerando um valor fixo para a gravidade em Lavras/MG, $g=9,78 m/s^2$, foram plotados gr\'{a}ficos que retratam a 
varia\c{c}\~{a}o do \^{a}ngulo $\theta$ para um dado intervalo de tempo, com coeficientes de atrito 0,5; 0,1 e 0,05 e fazendo $a=6,5$ cm 
e $b=4,5$ cm.

\indent Na figura 4, mostra-se que quanto maior o coeficiente de atrito, mais tempo ser\'{a} necess\'{a}rio para que ocorra a 
eleva\c{c}\~{a}o do eixo horizontal para a vertical, isto \'{e}, varia\c{c}\~{o}es no \^{a}ngulo $\theta$. Por exemplo, vemos que, a 
medida que o coeficiente de atrito diminui, o tempo de eleva\c{c}\~{a}o para um dado \^{a}ngulo aumenta de um fator de 5. J\'{a}, na 
figura 6, o fator de acr\'{e}scimo no tempo \'{e} de 10.\\
\indent No caso em que se utiliza o semieixo maior com $a=6,5$ cm na figura 4, comparado com o semieixo maior $a=5,5$ cm na figura 6, e mantendo 
fixo $b=4,5$ cm, percebe-se que quanto mais comprido o semieixo maior, menor ser\'{a} o tempo gasto para a eleva\c{c}\~{a}o  do eixo da 
horizontal para a vertical. Para o caso de $a=5,5$ cm, da figura 6, novamente reparamos que quanto menor o coeficiente de 
atrito\footnote{O coeficiente de atrito, para esse caso, n\~{a}o deve ser nulo.}, maior ser\'{a} o tempo de rota\c{c}\~{a}o do elipsoide.\\
\indent Se considerarmos o caso especial em que $a=b+\delta$, limite especial para esfera, $\delta\rightarrow0$ com $\delta\neq0$, teremos:

\begin{eqnarray}
 q&=&\frac{5g(a^{2}-b^{2})}{2(a^{2}+b^{2})}\Rightarrow q=\frac{5g(b^{2}+2b\delta+\delta^{2}-b^{2})}{2(b^{2}+2b\delta+\delta^{2}+b^{2})}  \nonumber \\
 q&=&\frac{5g}{2}\bigg(\frac{\delta}{b+\delta}\bigg)=\frac{5g}{2b}\bigg[1-\frac{\delta}{b}\bigg] \nonumber \\
 q&=&\frac{5g}{2b}\delta,\label{esfera}
\end{eqnarray}

\noindent onde $\delta$ \'e uma pequena varia\c{c}\~ao no raio da esfera e $\delta^{2}$ \'e desprez\'ivel. Substituindo a equa\c{c}\~ao 
dada por (\ref{esfera}) na equa\c{c}\~ao (\ref{30}), verifica-se que esse resultado n\~ao satisfaz a condi\c{c}\~{a}o de equil\'ibrio do 
girosc\'opio, pois se $\delta\rightarrow0$ teremos que $q\rightarrow0$, assim $\Omega^{2}\rightarrow0$. Quando $\delta\neq0$ e \'e um
valor pequeno, podemos analisar esse movimento por meio da an\'alise num\'erica da equa\c{c}\~ao dada abaixo:

\begin{equation}
 \tan{\theta}=e^{-\mu(5g/2b)\delta(t-t_{0})}. \nonumber
\end{equation}

\begin{figure}
 \begin{center}
 \includegraphics[scale=0.3,angle=0]{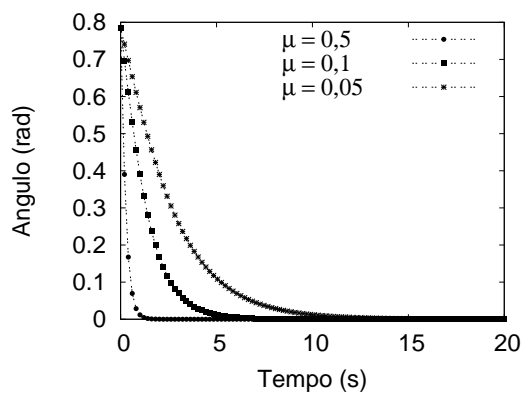}
\end{center}
\caption{Representa\c{c}\~{a}o da varia\c{c}\~{a}o do \^{a}ngulo de eleva\c{c}\~ao $\theta$, ao longo do tempo t, com as dimens\~{o}es espaciais de 6,5 cm e 4,5 cm. Sendo os valores de 0,5; 0,1 e 0,05 ilustrados por uma linha com pontos, uma linha com quadrados e uma linha com asteriscos, respectivamente.}
\end{figure}

\noindent S\~ao utilizadas as mesmas considera\c{c}\~oes feitas nas duas an\'alises num\'ericas acima, isto \'e, $g=9,78m/s^{2}$ e 
coeficientes de atrito 0,5; 0,1 e 0,05, como mostrado na figura 5. \

\begin{figure}
\begin{center}
  \includegraphics[scale=0.3,angle=0]{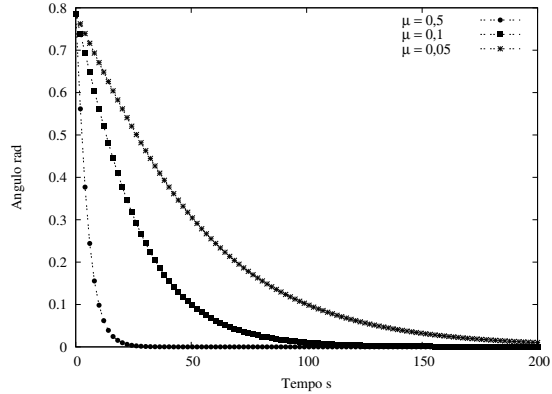}
\end{center}
\caption{Representa\c{c}\~{a}o da varia\c{c}\~{a}o do \^{a}ngulo de eleva\c{c}\~ao $\theta$ ao longo do tempo t, quando $a=b+\delta$ com 
$\delta\rightarrow0$, para o caso de uma esfera.Sendo os valores de 0,5; 0,1 e 0,05 ilustrados por uma linha com pontos, uma 
linha com quadrados e uma linha com asteriscos, respectivamente.}
\end{figure}

\indent Observa-se na figura 5, que ao se fazer a aproxima\c{c}\~ao para uma esfera, o tempo gasto para atingir o estado de 
eleva\c{c}\~{a}o do eixo horizontal para a vertical \'e bem maior quando comparado com o tempo gasto pelo elipsoide. E quanto menor o 
coeficiente de atrito, mais tempo \'e necess\'ario para que ocorra essa eleva\c{c}\~ao. Assim torna-se dif\'{i}cil visualizar esse 
movimento em um experimento real, visto que a eleva\c{c}\~ao em si, \'e r\'apida e tamb\'em pela pr\'opria simetria da esfera.

\newpage

\section*{An\'{a}lises e Conclus\~{o}es}
O fen\^{o}meno discutido neste artigo foi foco de estudos durante um longo tempo, sendo encontrada a solu\c{c}\~{a}o apenas em 2002, a 
partir das leis de Newton. Neste trabalho, foram utilizadas as fun\c{c}\~oes de Lagrange, cujo lagrangiana do sistema \'{e} dado pela 
equa\c{c}\~{a}o (\ref{16}), para determinar as equa\c{c}\~{o}es de movimento, as quais conferem com as encontradas por Moffatt e Shimomura.
 Ao se encontrar as equa\c{c}\~{o}es de movimento, estudamos um caso particular, onde a integra\c{c}\~{a}o do sistema s\'{o} foi 
poss\'{i}vel em decorr\^encia do invariante de Jellett.

\begin{figure}
\begin{center}
  \includegraphics[scale=0.3,angle=0]{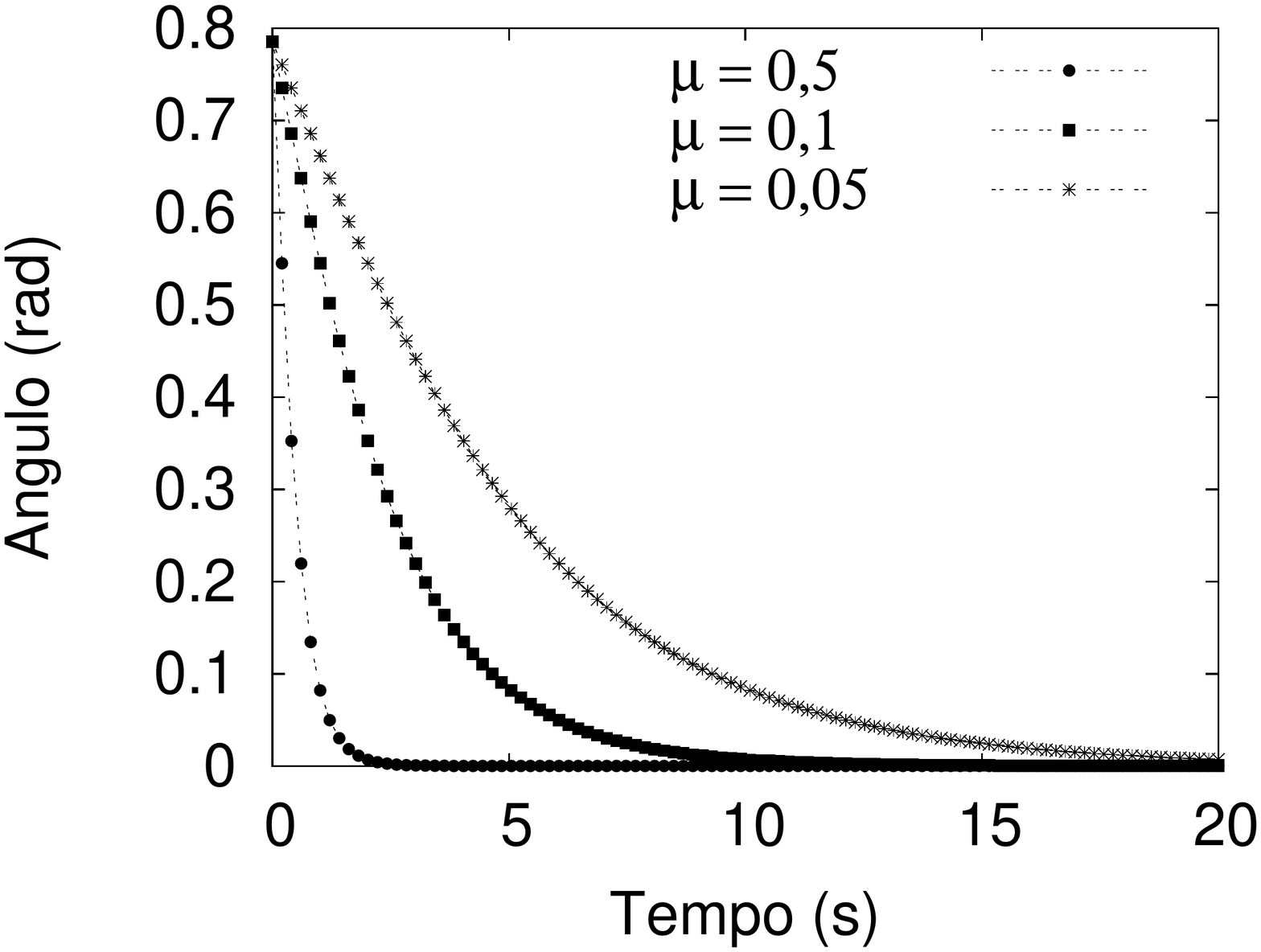}
\caption{Representa\c{c}\~{a}o da varia\c{c}\~{a}o do \^{a}ngulo de eleva\c{c}\~ao $\theta$ ao longo do tempo t, com as dimens\~{o}es 
espaciais de 5,5 cm e 4,5 cm.Sendo os valores de 0,5; 0,1 e 0,05 ilustrados por uma linha com pontos, uma linha com quadrados e uma linha 
com asteriscos, respectivamente.}
\end{center}
\label{fig:imagem2}
\end{figure}

\newpage

\indent A solu\c{c}\~{a}o encontrada descreve a eleva\c{c}\~{a}o do eixo horizontal para a vertical, durante o movimento de precess\~{a}o, 
com a condi\c{c}\~{a}o de equil\'{i}brio sendo satifeita. Por exemplo, para um ovo cru essa condi\c{c}\~ao de equil\'ibrio n\~ao pode ser 
satisfeita, pois a energia transmitida para girar o ovo \'{e} dissipada pelo seu fluido interior.\\
\indent Para observar de forma qualitativa o fen\^{o}meno descrito nessa teoria, foi feito um experimento demonstrativo, onde utilizamos 
diferentes coeficientes de atrito, tais como: uma superf\'{i}cie com atrito desprez\'ivel, mas n\~ao nulo, e superf\'{i}cies rugosas, 
como por exempo, folhas de lixa. Podendo observar experimentalmente que quanto maior o coeficiente de atrito, menor o tempo de 
eleva\c{c}\~{a}o do eixo horizontal para a vertical, como descrito na teoria. Acesse o link [4] e assista ao v\'{i}deo.\\
\indent Os m\'etodos e resultados deste trabalho podem ser utilizados numa disciplina de mec\^anica te\'orica, quando se estuda o t\'opico 
da din\^amica de um corpo r\'igido.

\section*{Agradecimento}
Agradecemos a Prof. Dra. Karen Luz Burgoa Rosso, do Departamento de Ci\^{e}ncias Exatas da Universidade Federal de Lavras pela gentileza 
de realizar e orientar as contru\c{c}\~{o}es das figuras presentes neste artigo, al\'{e}m de ler todo o trabalho e fazer as 
corre\c{c}\~{o}es necess\'{a}rias e tamb\'{e}m pelas observa\c{c}\~{o}es de ordem est\'{e}tica.

\section*{Ap\^{e}ndice}
Para descrever a din\^{a}mica de um corpo r\'{i}gido qualquer, pode-se trabalhar com os \^{a}ngulos de Euler 
$\theta,\psi,\varphi$ \cite{9}. Considere a figura abaixo:

\begin{figure}[!htb]
\centering
\includegraphics[scale=0.29,angle=0]{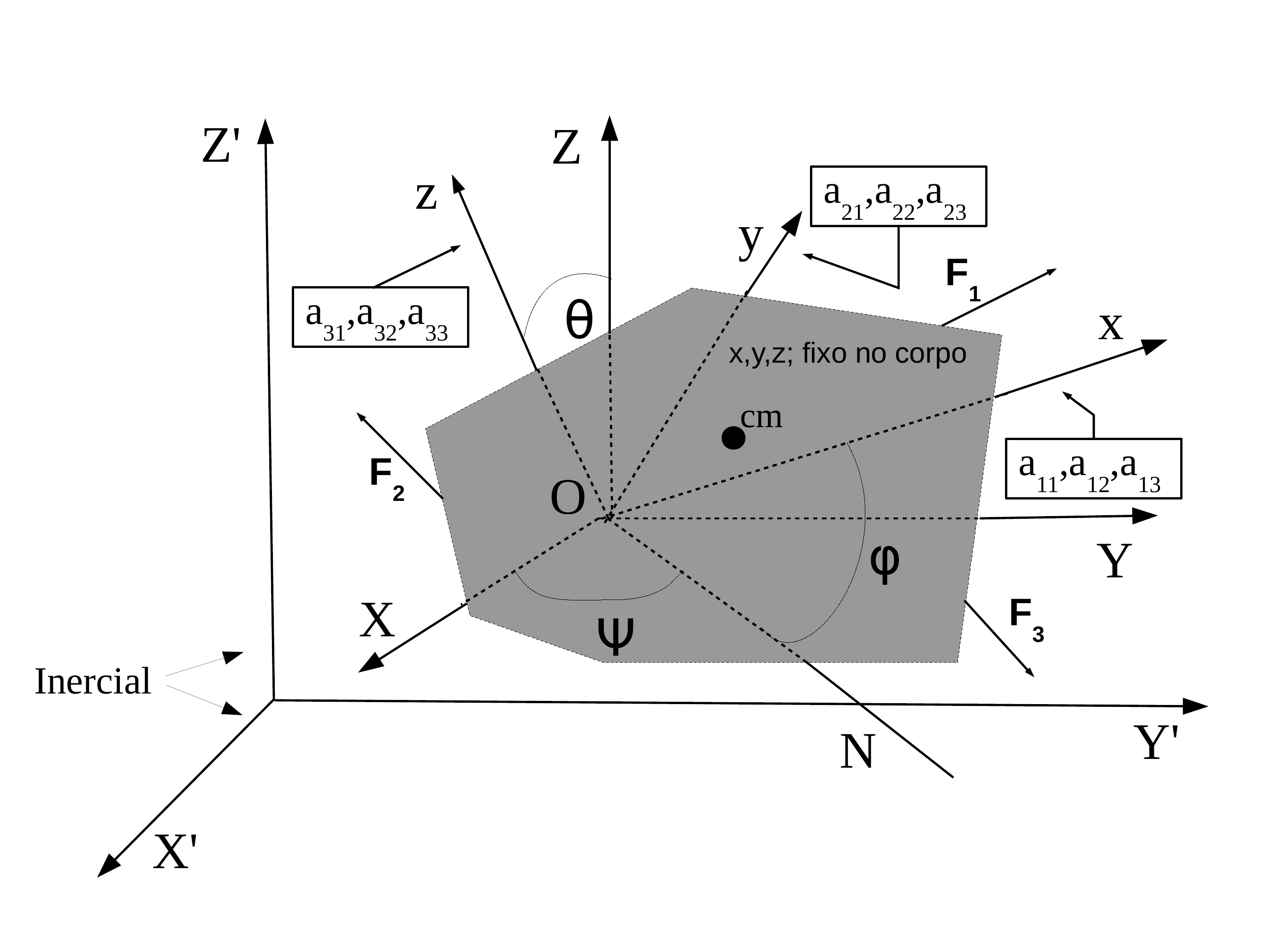} 
\caption{Esse corpo r\'{i}gido realiza um movimento decorrente da a\c{c}\~{a}o de for\c{c}as $\bf F_{1}$, $\bf F_{2}$ e $\bf F_{3}$. O sistema 
de coordenadas Oxyz est\'{a} fixo no corpo, sendo que a origem \'{e} coincidente com a origem do sistema OXYZ, o qual \'{e} paralelo ao 
sistema inercial OX'Y'Z'. Sejam $a_{11}$, $a_{12}$, $a_{13}$ os cossenos diretores de Ox relativos a OXYZ.}
\end{figure}

O sistema de coordenadas OXYZ \'{e} fixo no espa\c{c}o e o sistema Oxyz, fixo no corpo. A rela\c{c}\~{a}o entre esses dois sistemas de 
coordenadas, envolvendo os \^{a}ngulos de Euler, \'{e} dada pala matriz de transforma\c{c}\~ao $\bf{A}$:
\[ \bf{A}=\left( \begin{array}{ccc}
     a_{11} & a_{12} & a_{13} \\
     a_{21} & a_{22} & a_{23} \\
     a_{31} & a_{32} & a_{33}
          \end{array}\right)\Rightarrow\]

\[ \bf{A}=\left( \begin{array}{ccc}
\cos\psi\cos\varphi - \sin\psi\sin\psi\cos\theta & \cos\psi\sin\varphi + \sin\psi\cos\varphi\cos\theta & \sin\theta\sin\psi \\
- \sin\psi\cos\varphi - \cos\psi\sin\varphi\cos\theta & - \sin\psi\sin\varphi + \cos\psi\cos\varphi\cos\theta & \sin\theta\cos\psi \\
\sin\theta\sin\varphi & -\sin\theta\cos\varphi & \cos\theta \end{array} \right)\] 

Considere que as componentes $f'_{xi}$, $f'_{yi}$, $f'_{zi}$ de $F_{i}$, sejam paralelas a X,Y,Z e as coordenadas 
$x_{i}$, $y_{i}$, $z_{i}$ dos pontos de aplica\c{c}\~{a}o de $p_{i}$ relativos a x, y, z, sejam conhecidos. As for\c{c}as generalizadas 
correspondentes para $x_{1}$, $y_{1}$, $z_{1}$, s\~{a}o:

\begin{center}
$F_{x1}= \sum f'_{xi}  \quad\mbox,\quad
F_{y1}= \sum f'_{yi}  \quad\mbox{e}\quad 
F_{z1}= \sum f'_{zi}  $
\end{center}

Para descobrir $F_{\theta}$, por exemplo, procede-se da maneira a seguir. A componente do vetor torque total, ao redor do eixo x, \'{e} 
dado por $\tau _{x} = \sum (f_{zi}y_{i}-f_{yi}z_{i})$, onde $f_{xi}$ \'{e} a componente de $F_{i}$ na dire\c{c}\~{a}o do eixo X do 
sistema de coordenadas fixo no corpo. Mas $f_{xi}=f'_{xi}\alpha_{11}+f'_{yi}\alpha_{12}+f'_{zi}\alpha_{13}$, ... para $f_{yi}$ e $f_{zi}$. 
Ap\'{o}s determinado $\tau_{x}$, $\tau_{y}$, $\tau_{z}$, encontra-se $\tau_{\theta}=\tau_{x}\cos\psi-\tau_{y}\sin\psi=F_{\theta}$. 
Pode-se encontrar express\~{o}es similares para $\tau_{\varphi}$ e $\tau_{\psi}$, tal que:

\begin{center}
$F_{\theta}=\tau_{x}\cos\psi-\tau_{y}\sin\psi  \quad\mbox,\quad
F_{\psi}=F_{z}$ 
\end{center}
\be
F_{\varphi}=\tau_{x}\sin\theta\sin\psi+\tau_{y}\sin\theta\cos\psi+\tau_{z}\cos\theta
\ee

Para calcular $\tau_{x}$, $\tau_{y}$, $\tau_{z}$, faz-se:

\begin{eqnarray}
  {\bf {\tau}}={\bf {X_{p}}} \wedge {\bf {F'}}= (X_{p},0,Z_{p})\wedge (F_{x},F_{y},R) \nonumber \\
=-Z_{p}F_{y} {\bf {I}}+ (Z_{p}F_{x}-X_{p}R) {\bf {J}} + X_{p}F_{y} {\bf {K}}
\end{eqnarray}

\noindent onde $X_{p}$ \'{e} o vetor posi\c{c}\~{a}o que localiza o ponto P e $\bf F'$ possui as componentes das for\c{c}as que atuam no corpo. Utilizando (1) e (5), pode-se escrever:

\be
  {\bf {\tau}}=-z_{p}F_{y} {\bf {i}}+ (Z_{p}F_{x}-X_{p}R) {\bf {j}} + x_{p}F_{y} {\bf {k}}. \label{33}
\ee

Neste artigo, foi enfatizado o comportamento da din\^{a}mica para o \^{a}ngulo $\theta$, considere ent\~{a}o $\psi = \frac{3\pi}{2}$. Sustituindo o valor de \^{a}ngulo $\psi$ e (\ref{33}), t\^{e}m-se:

\begin{center}
$F_{\theta}=\tau_{y}=Z_{p}F_{x}-X_{p}R  \quad\mbox,\quad
F_{\psi}=\tau_{z}=x_{p}F_{y}$ 
\end{center}
\begin{center}
$F_{\varphi}=-\tau_{x}\sin\theta+ \tau_{z}\cos\theta = -z_{p}F_{y}\sin\theta + x_{p}F_{y}\cos\theta$
\end{center}

Portanto, $F_{\theta}$, $F_{\psi}$, $F_{\varphi}$ s\~{a}o as for\c{c}as generalizadas.


\end{document}